\documentclass[fleqn,12pt]{wlscirep}
\usepackage[utf8]{inputenc}
\usepackage[T1]{fontenc}
\usepackage{lineno}
\usepackage{soul}

\newcommand{\beq}{\begin{equation}}
\newcommand{\eeq}{\end{equation}}
\newcommand{\bea}{\begin{eqnarray}}
\newcommand{\eea}{\end{eqnarray}}
\newcommand{\be}{\begin{equation}}
\newcommand{\ee}{\end{equation}}

\newcommand{\eq}[1]{Eq.~\ref{#1}}
\newcommand{\fig}[1]{Fig.~\ref{#1}}
\newcommand{\tab}[1]{Tab.~\ref{#1}}
\newcommand{\sect}[1]{Sect.~\ref{#1}}

\newcommand{\coo}{CO$_2$ }

\usepackage[normalem]{ulem}

\colorlet{mgreen}{green!50!black!50!}

\usepackage{multirow}

\title{Maritime transportation and people mobility in the early diffusion of COVID-19 in Croatia}

\author[1,*]{Corentin Cot}
\author[2]{Dea Aksentijevi{\'c}}
\author[2]{Alen Jugovi{\'c}}
\author[3]{Giacomo Cacciapaglia}
\author[4]{Gianandrea Mannarini}

\affil[1]{Laboratoire de Physique des 2 Infinis Ir\`ene Joliot Curie (UMR 9012), CNRS/IN2P3, 15 Rue Georges Clemenceau, 91400 Orsay, France}
\affil[2]{Pomorski fakultet Sveučilišta u Rijeci / Faculty of Maritime Studies, University of Rijeka, Studentska 2,
51000 Rijeka, Croatia}
\affil[3]{Univ Lyon, Univ Claude Bernard Lyon 1, CNRS/IN2P3, IP2I Lyon, UMR 5822, F-69622, Villeurbanne, France}
\affil[4]{Fondazione CMCC – Ocean Predictions and Applications Division, via Marco Biagi 5, 73100 Lecce, Italy}

\affil[*]{corresponding author: corentin.cot@ijclab.in2p3.fr}

\begin{abstract}
\noindent
The outbreak of COVID-19 in Europe occurred in early 2020. During the year, several waves of infection developed with different timings across the European countries. The onset of the largest wave of infection occurred in August-September.  Croatia is a hotspot of tourism in the Mediterranean region, and thus may have acted as an incubator of the pandemic during the summer of 2020. Given this, we designed a data-driven investigation to assess the possible role of mobility of passengers to and within Croatia through various modes of transportation. To this end, observational datasets  were integrated with the modelling framework of the ``epidemic Renormalisation Group''. Comparing the models to the epidemiological data allowed to disfavour, in the case of Croatia in 2020, any prominent role in propagating the infection by either maritime or train transportation, while highlighting the leading role of both road and airborne mobility. The proposed framework aims to test hypotheses regarding the causation of infectious waves, with the capacity to rule out unrelated phenomena.
\end{abstract}
\begin{document}

\flushbottom
\maketitle
\thispagestyle{empty}

\section{Introduction}

With the growth of human population and  its impact on the environment, our societies are becoming increasingly vulnerable to new diseases, especially viral infectious diseases of zoonotic origin. At present, just 3\% of the land ecosystems are untouched by human activities \cite{10.3389/ffgc.2021.626635}.  Furthermore, human-induced climate change is causing relocation of species and rapid migration of humans, hence increasing cross-species viral transmission risks \cite{climatechange}. The indirect impact caused by the thawing of the Arctic permafrost also poses the risk of releasing past viral charges \cite{permafrost}. In addition to this, the economic globalisation has increased the mobility of both goods and people across countries and continents, hence facilitating the global spread of disease carriers. All these factors contribute to the transmission of viral pathogens from animal species to humans, and their rapid diffusion within the world population. The COVID-19 pandemic \cite{COVID19-1,COVID19-2} showcased this process \cite{Bluedot}. It also dramatically showed the unpreparedness of human society to face the threat of a pandemic \cite{preparedness1} and its inability to efficiently cope with the effects, evident in the emergence of multiple epidemiological waves \cite{Lancet}. 

Hence, it has become of paramount importance to define and introduce protocols and preparedness measures that help governments, private companies, and individual citizens to face a new viral pandemic in its early phases. In this context, human mobility plays a crucial role in determining the transmission of the pathogens,  highlighting the importance of restriction measures at the beginning of a pandemic\cite{Chinazzi,Lai2020,Flaxman2020,SEIR,scala2020}. Examples of these restriction measures  include lockdowns, mobility limitations within countries, and border closures. Travel between countries and geographical regions, in fact, may have played a crucial role in the diffusion of epidemic waves within and among continents, e.g. causing multiple waves in Europe right after the lockdowns were lifted \cite{Cacciapaglia:2020mjf,cacciapaglia2020second}. This effect was not widely expected in the scientific community, as diverse scenarios for the short and long term evolution of the pandemic were on the table \cite{Scudellari}.  Also, it was found that  diffusion of the infection within a community was mainly driven by specific {\it superspreader locations} \cite{Leskovec2020}, where more intense social interactions occur.

The main objective of this paper is to quantitatively study the impact of various passenger transportation vectors on the early diffusion of COVID-19. As a case study,  Croatia was we considered, with a special focus on the impact of maritime transportation. It has been shown that maritime transportation has been crucially affected by the COVID-19 pandemic \cite{March_2021_nature}, with an impact also on its greenhouse gas emission \cite{mannarini_su2022}. The initial cases in In Croatia were reported in March-April. Later on, both a second and third wave of infections occurred between June and September 2020. As shown in Fig.~\ref{fig:CIMIS-COVID}, this coincided with the reopening of the sea-based touristic links. 

Maritime passenger transportation is crucial for Croatia due to its long and archipelagic coast with more than one thousand islands on the Adriatic Sea.  Regular ferry  traffic between Croatia and Italy is extremely significant, which takes place via the ports of Split, Zadar and Dubrovnik \cite{dundovic2012analysis}, connecting them with the Italian ports of Ancona and Bari. Many tourists reach the Croatian coasts by this means. Fr this reason, ships may have played the role of a superspreader, triggering the pandemic wave that hit Europe in the summer of 2020. It is worth noting that the situation in Croatia in June-August of 2020 is of particular interest, as it showed an earlier increase of infections after the lockdowns were lifted, as compared to other European countries \cite{cacciapaglia2020second}. We remark that passenger mobility plays a crucial role in the diffusion of infections in the early phases of a pandemic, while at later stages variants due to genetic mutations start became the predominant factor in the emergence of new epidemic waves \cite{cacciapaglia2022earlywarning}.

This study was devised and conducted in the frame of the 
GUTTA project\cite{webGUTTA}, part of the Italy-Croatia Interreg Programme. GUTTA’s main goal  was the development and public release of a  tool, 
GUTTA-VISIR \cite{webGUTTAVISIR}, for operational provision of least-\coo ferry routes, based on forecast marine conditions in the Adriatic and Northern Ionian seas \cite{mannarini_jmse2021}. However, the project also investigated the \coo emissions from ferries in Europe, including an assessment of how the COVID-19 pandemic hit the sector \cite{mannarini_su2022}.

To reach the main objective of this study, we performed a data-driven analysis that quantifies the impact on the diffusion of COVID-19 of both terrestrial and maritime passenger transportation between Croatia and its neighbouring countries. The role, if any, of maritime transportation is not easily assessed ex-ante. On the one hand, the amount of passengers carried by sea was smaller than compared to other modes of transportation. The data shows that maritime passengers have been less than 1\% of the car passengers entering and leaving Croatia in 2020. On the other hand, ferry  passengers were transported in limited volumes for relatively long times,  which might have favoured the diffusion of the virus \cite{Mizumoto_2020}.  The latter hypothesis would also be suggested by the timing of the infection diffusion across summer 2020, coinciding with the reopening of ferry connections across the Adriatic.  

The datasets used in this study included the passenger flow to Croatia via waterborne, airborne, and terrestrial transportation modes. To connect mobility to the epidemiological data, we employed a novel approach to infectious disease spreading, the epidemiological Renormalisation Group framework (eRG) \cite{DellaMorte:2020wlc}, which is inspired by theoretical high energy physics \cite{Wilson:1971bg,Wilson:1971dh}. The eRG offers a computationally inexpensive characterisation of a single wave diffusion in terms of just two constant parameters. Once extended to a network of semi-isolated populations \cite{Cacciapaglia:2020mjf}, it will enable to study the spread of the infectious disease among regions/Countries. This method was pivotal in predicting the 2020 second wave in Europe \cite{cacciapaglia2020second}, and has been tested in the US as well \cite{cacciapaglia2020better}. The main advantage of the eRG approach is the ability to characterise a wave in all its phases, from the initial exponential increase to the peak and reduction of the new infections, in terms of a simple logistic function. It is, however, not well suited for short-term forecasting \cite{Perc2020}, for which more traditional compartmental models are preferable \cite{Kermack:1927}. It is noteworthy that the eRG solutions are related to the simplest compartmental model, based on Susceptible-Infectious-Removed (SIR), with time-dependent parameters \cite{Della_Morte_2021}. For our purpose, the eRG offers a reliable handle to quantify the impact of mobility on the timing of the peak of the third wave in different regions of Croatia.

The remainder of this paper will outline the methodology, in \sect{sec:Methodology}, and the results, in \sect{sec:Results}. Conclusions and recommendations will be described  in \sect{sec:Conclusions}.

\section{Methodology}\label{sec:Methodology}

\begin{figure}[t]
\centering
\includegraphics[width=13.5cm]{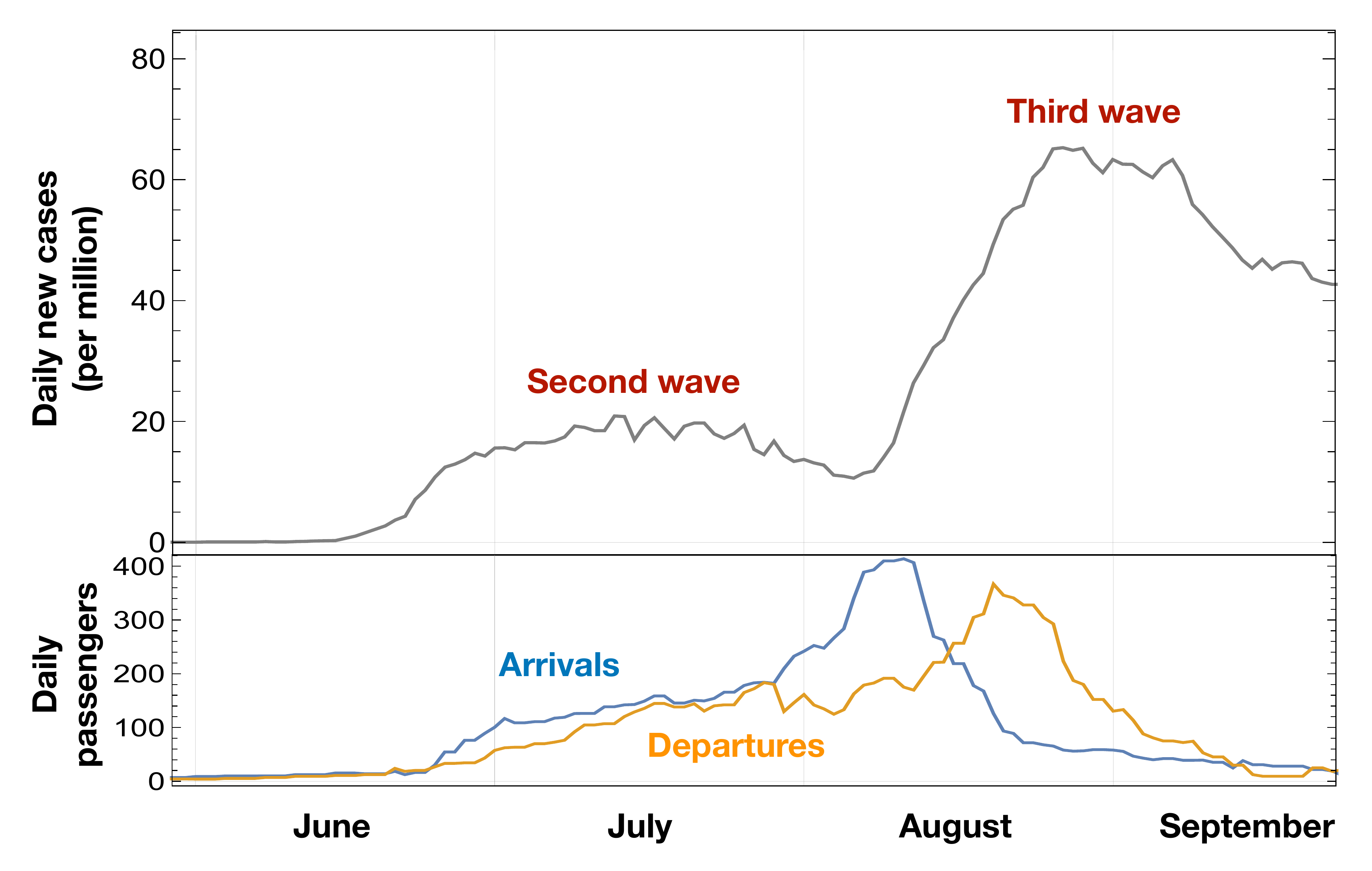}  
\caption{COVID-19 new cases (grey line) in Croatia between June and September, 2020, shown as weekly averaged data. As a comparison, we report the number of maritime passengers arriving or departing from Croatian ports in the same period. } \label{fig:CIMIS-COVID} 
\end{figure}

To study the diffusion of the COVID-19 infections in Croatia, we combined data describing people’s mobility via various transportation modes with an epidemiological dataset. The latter consists of the daily number of newly infected individuals that were tested positive, during a period of time, in each county\cite{webNUTS} of Croatia. The mobility data comprises the number of individuals entering Croatia via sea, land or air, provided by various sources and described in detail below. The correlation between the two datasets was studied within the eRG framework, consisting of a set of coupled differential equations \cite{Cacciapaglia:2020mjf}. The main advantage of the mathematical model provided by the eRG is to allow characterisation. This includes the diffusion of an infectious disease within connected regions. Hence, the position of the peaks, i.e., the timing of the local maxima of new infections in different regions, can be predicted as a function of the mobility data. Comparing the predictions of the model with the actual data allows us to determine the role of various transportation modes in facilitating the diffusion of COVID-19. This mechanism is expected to be the dominant mechanism of diffusion at the beginning of the pandemic. In Croatia, the first epidemiological wave, characterised by a temporary exponential increase of the cases, took place in March through April 2020. After a period when the rate of infections slowed down due to the lockdowns, a new increase was detected starting towards the end of June and lasting through the end of July, followed by another increase in August and September. We identify these two episodes as the ‘second’ and ‘third waves’ respectively, see Fig.\ref{fig:CIMIS-COVID}.

The rest of this section provides: a detailed presentation of the datasets used for this research in \sect{sec:datasets}, the procedure for a geographical aggregation of the data, outlined in \sect{sec:NUTS}, and the application of the eRG framework to this specific data in \sect{sec:eRG}.

\subsection{Datasets} \label{sec:datasets}

\begin{table}[]
\centering
\begin{tabular}{l|l|l|l|ll|c|}
\cline{2-7}
\multirow{2}{*}{}                                  & \multicolumn{1}{c|}{\multirow{2}{*}{Vehicles}} & \multicolumn{1}{c|}{\multirow{2}{*}{Name}} & \multicolumn{1}{c|}{\multirow{2}{*}{Provider}} & \multicolumn{2}{c|}{Resolution}                           & \multirow{2}{*}{\#datapoints} \\ \cline{5-6}
                                                   & \multicolumn{1}{c|}{}                          & \multicolumn{1}{c|}{}                      & \multicolumn{1}{c|}{}                          & \multicolumn{1}{c|}{Time}    & \multicolumn{1}{c|}{Space} &                              \\ \hline
\multicolumn{1}{|l|}{Maritime}                     & Ferries  (Ro-Pax)                                       & CIMIS                                      & MMPI                                           & \multicolumn{1}{l|}{daily}  & by port                    & 1360                         \\ \hline
\multicolumn{1}{|l|}{\multirow{2}{*}{Terrestrial}} & Cars                                           & Highway data                               & UNIRI                                          & \multicolumn{1}{l|}{summer}  & borders                    & 108                             \\ \cline{2-7} 
\multicolumn{1}{|l|}{}                             & Trains                                         & Railway data                               & UNIRI                                          & \multicolumn{1}{l|}{annual}  & borders                    & 14                             \\ \hline
\multicolumn{1}{|l|}{Airborne}                     & Planes                                         & Air traffic data                           & MMPI                                           & \multicolumn{1}{l|}{monthly} & airports                   & 21                           \\ \hline
\multicolumn{1}{|l|}{Epidemiology}                 & —                                              & New cases                                  & MMPI                                           & \multicolumn{1}{l|}{daily}   & county                     & 575                          \\ \hline
\end{tabular}
\caption{Description of mobility datasets used in this work.}\label{tab:01}
\end{table}

The following datasets for 2020 (see summary in Table~\ref{tab:01}) were used for the numerical analyses:

\begin{itemize}
\item[1)] Maritime transportation data was obtained from the {\it Croatian Integrated Maritime Information System} (CIMIS). The dataset, provided by the GUTTA partner ``{\it Ministarstvo Mora, Prometa i Infrastrukture}'' (MMPI -- Ministry of Maritime Affairs, Transport and Infrastructure of Croatia), provides information for each Croatia seaport regarding departures and arrival times of ferries along with the number of both embarking and disembarking passengers. We included data from car-passenger ferry routes, which include the routes Ancona-Zadar, Ancona-Split and Bari-Dubrovnik. 

\item[2)]	Car traffic data was collected by the {\it University of Rijeka} (UNIRI) by contacting the limited liability company ``Hrvatske Ceste'' \cite{webHC}, which has a function of management, construction and maintenance of state roads.  
This dataset provides the number of travellers crossing each Croatian border control checkpoint per year. Hence, we reconstructed the average flow between the neighbouring countries (Slovenia, Hungary, Serbia, Bosnia-Herzegovina, Montenegro) and the considered Croatian regions, both entering and leaving the country by road. The data also contains the number of travellers passing various checkpoints along the major Croatian roads inside the country, however this information was discarded as it did not allow to reliably reconstruct people's mobility within Croatia.

\item[3)] Railway traffic data was extracted from the ``{\it Independent Regulators' Group}'' IRG-rail 2021 report \cite{webIRGrail}, provided by MMPI and from {\it HŽ Infrastruktura}. The latter organisation is responsible for the railway system in Croatia. The 2021 report also outlines the impact of the COVID-19 crisis on the network during the first half of 2020. Data of network topology, passenger flows, and operational conditions was taken from the report.  The railway traffic was recorded on an annual basis. 

\item[4)] Air traffic data was procured by the MMPI from the {\it Eurocontrol Air Traffic Directorate} in Lyon, France. The dataset consisted of the number of travellers per month per airport in Croatia for 2020.

\item[5)] The COVID-19 epidemiological data was extracted from an open-source Croatian public resource, ``koronavirus.hr'' \cite{webKorona}. The dataset includes the cumulative total number of infections and the number of daily new infected individuals for each Croatian county. A new infection is counted for each individual that reported a new positive test. We extracted the data from the 21${}^{\rm st}$ of March, 2020, to the 18${}^{\rm th}$ of October, 2021, corresponding to 575 days in total. The raw data was pre-processed to smooth daily fluctuations by applying a moving 7-day averaging procedure.
\end{itemize}

\begin{figure}[t]
\centering
\includegraphics[width=17cm]{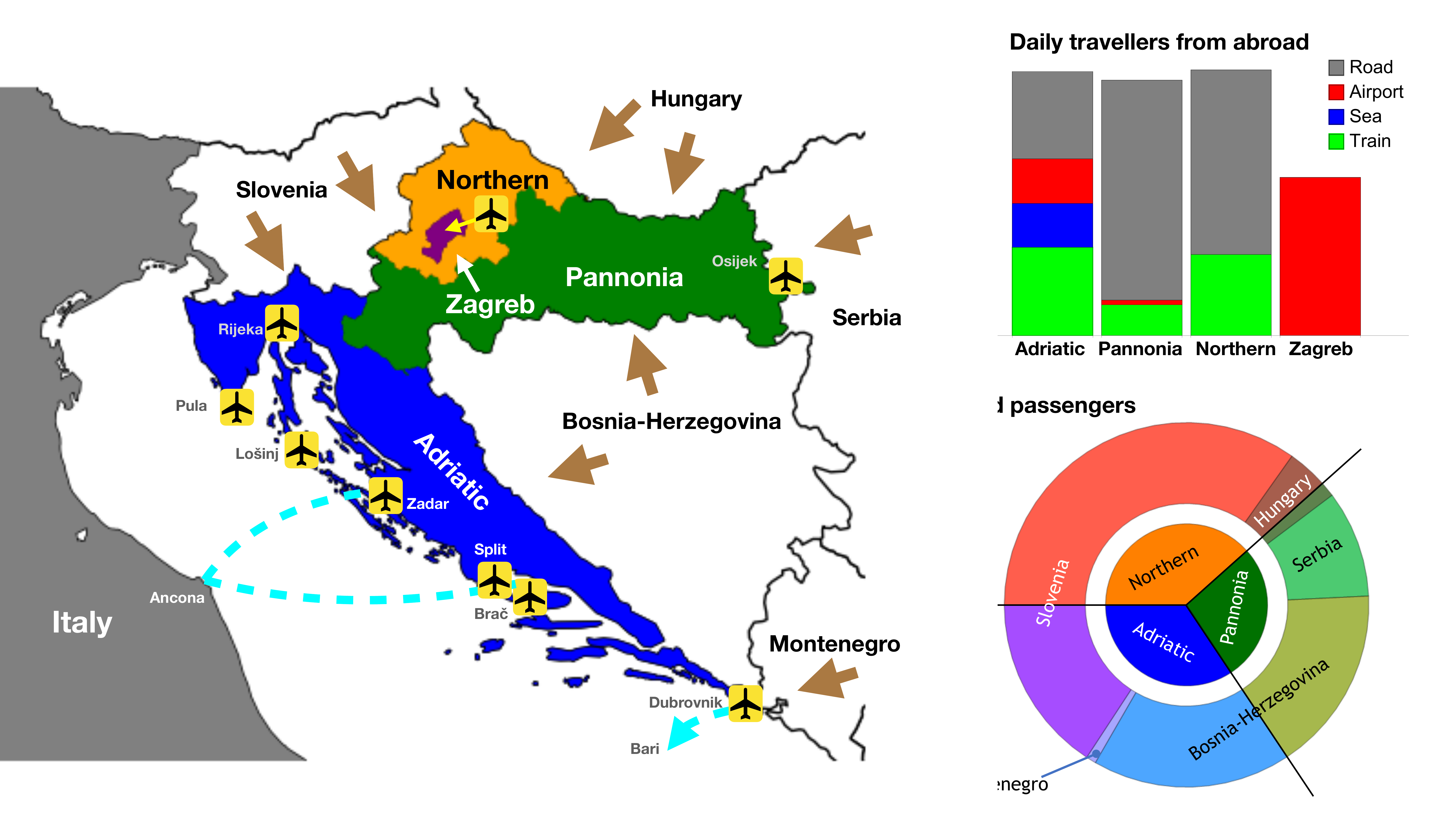}  
\caption{{\bf A)} NUTS-2 regions of Croatia, indicating the main mobility axes via land (brown arrows), flights (international airports at Zagreb, Osijek, Rijeka, Pula, Lošinj, Zadar, Split, Brač and Dubrovnik) and sea (car-passenger ferry lines in cyan). {\bf B)} Partition of the passenger inflow from abroad to the four NUTS-2 regions. The histograms are stacked in log scale, showing a clear dominance of road passenger flow.   {\bf C)} Partition of the road passenger flow among Croatian regions and the neighbouring countries. The outer ring shows the flow from each country, proportionally distributed to the regions. \label{fig:map-histo} }
\end{figure}

We offer a visualisation of the mobility data in Fig.~\ref{fig:map-histo}, subdivided according to the regions of Croatia (``Pannonia'', ``Adriatic'', ``Northern'', ``Zagreb'') we define in the next section. In particular, in Fig.~\ref{fig:map-histo}B we show the average number of  daily passengers entering each region from abroad, plotted as stacked histograms. This proves that the dominant flow is due to cars, followed - one order of magnitude below - by the airborne traffic. The only exception is Zagreb, which is an enclosed region, hence its only  direct connection with other countries is through airborne transport. The maritime passenger flow is only relevant for the Adriatic region, where it only constitutes roughly 1\% of the total.
Finally, Fig.~\ref{fig:map-histo}C shows a subdivision of the car passengers by region, as shown in the inner circle, and by neighbouring country, as shown in the outer ring.  Note that the angle of each wedge is proportional to the number of daily passengers. The inner circle visualises the proportion of the total passengers entering Croatia from the five neighbouring countries, divided among the three relevant Croatian regions (Pannonia, Northern and Adriatic). The ring shows the proportion of such passengers from each bordering country, highlighting the proportional distribution of passengers from Slovenia to Northern and Adriatic (no significant flow to Northern is observed in the data), from Hungary to Northern and Pannonia, and finally from Bosnia-Herzegovina to Pannonia and Adriatic.

\subsection{NUTS-2 regions and how to deal with Zagreb} \label{sec:NUTS}
Croatia is a diverse country in terms of geography, population density, and demography. It consists of an elongated coastal region of great touristic interest, and an internal region where the capital city Zagreb is located. Furthermore, it borders with numerous countries: Slovenia and Hungary in the North, Serbia in the East, Bosnia-Herzegovina and Montenegro in the South. It also shares a maritime border with Italy in the west (across the Adriatic Sea). The main connections with abroad, therefore, are realised via road and railway, via ferries (with Italy mainly) and flights. 

To characterise the diffusion of the virus within Croatia, we need to first establish regions within Croatia to be associated to the eRG equations. One possibility is to consider the Croatian counties, taking into account the subtlety of the mobility within counties and with neighbouring countries. This corresponds, in the Eurostat nomenclature \cite{webNUTS}, to the “NUTS-3” level. However, for our purposes, this subdivision level would be problematic for multiple reasons. Firstly, the population of different counties is highly unequal, ranging from 50,000 inhabitants in {\it Lika-Senj} to 800,000 in {\it Grad Zagreb}. This unequal distribution of the population would amplify, for counties with low population, any statistical fluctuations in the epidemiological data. Furthermore, the unequal weight of different counties can bias the numerical output of the eRG computations. Secondly, it is challenging to quantify the mobility flows between counties due to the small size of these territorial units. Access to the flow data along main highways was provided  by the MMPI. However, this data would not be sufficient to provide a reliable estimate at county level. In fact, drivers passing through the checkpoints on the main roads may drive across several counties, where possible contacts with infected individuals could take place. As a consequence, we deemed this level of geographical granularity to be inappropriate for our purposes. 

We instead opted for the 2021 NUTS-2 level, where Croatia is subdivided into four statistical regions as shown in Fig.~\ref{fig:map-histo}A:

\begin{itemize}
\item	Pannonia (HR02, {\it Panonska Hrvatska}): $1,054,000$ inhabitants, stretching to the East and adjoining Hungary, Serbia and Bosnia-Herzegovina.
\item	Adriatic (HR03, {\it Jadranska Hrvatska}): $1,372,000$ inhabitants, comprising of the coastal region and bordering Slovenia, Bosnia-Herzegovina and Montenegro, including Italy via sea.
\item	Northern (HR06, {\it Sjeverna Hrvatska}): $813,000$ inhabitants, in the north and bordering Slovenia and Hungary.
\item	Zagreb (HR05, {\it Grad Zagreb}): $800,000$ inhabitants, enclosed within the Northern region.
\end{itemize}

\noindent This division is much more uniform in terms of population, hence minimising the statistical uncertainties of the data. It is also more suitable for studying the diffusion of the infectious disease, as each region has its own specificity that makes its role unique. For instance, Adriatic is the only region that is connected by maritime transportation as it encompasses all the ports of Croatia. Furthermore, the main international airports are in Adriatic (Split) and Zagreb. All regions except Zagreb are connected to neighbouring countries via border road and train connections. 

The above information highlights an issue with the mobility data related to Zagreb: the absence of a region border with abroad. However, the traffic to the capital city is expected to be of major importance. To consider this missing information, in the numerical results, we ``re-routed'' part of the road and train flow across boundaries to the Zagreb region. By this it is meant that e.g. part of the passengers travelling via car (or train) from abroad to Pannonia (or Adriatic, or Northern) are supposed to end up their journey not in Pannonia but in the Zagreb region.
This is justified by the fact that highways and railway lines connect the boundaries directly to Zagreb, while crossing any of the other three NUTS-2 regions of Croatia.  We will investigate a few scenarios where a fixed fraction of the road and railway traffic from the surrounding regions is attributed to Zagreb.

\subsection{Applying the epidemic Renormalisation Group framework} \label{sec:eRG}

The eRG framework \cite{DellaMorte:2020wlc} was formulated as a simple mathematical tool to describe the exponential increase in the number of new infections, followed by a reduction back to approximately zero. We will refer to this phenomenon as a “wave”. As seen from Fig.~\ref{fig:CIMIS-COVID}, Croatia experienced three waves of COVID-19  between March and September 2020. More waves followed, showing larger numbers of infected individuals. It should be remarked, however, that the absolute number of cases in each wave cannot be completely trusted, as it depends on both the number of people that are subjected to tests and on biases in the testing strategies (e.g., correlations with hospitalisations, presence of asymptomatic cases, etc.).  

For a single isolated region, with constant population during the spread of a single wave, the eRG framework provides a first-order differential equation to describe the time-evolution of the number of individuals that contracted the disease. The eRG equation reads
\begin{equation} \label{eq:eRG}
    \frac{d \alpha}{d t} = \gamma\ \alpha \left( 1 - \frac{\alpha}{A} \right)\,,
\end{equation}
where $t$ is time and $\alpha$ is a non-dimensional function of the cumulative number of infected individuals in the region, $I_c (t)$. Hence, $\alpha (t)$ is a function of time only, where the spatial dependence has been integrated in. In principle, $\alpha$ can be any monotonic function of $I_c$, however comparison with data for COVID-19 and SARS showed that an optimal fit can be obtained for the natural logarithm \cite{DellaMorte:2020wlc}
\begin{equation}
    \alpha (t) = \ln \frac{I_c (t)}{N_m} \equiv \ln I_n (t) \,,
\end{equation}
where we normalised the number of infections by the population of the region in millions, $N_m$. Hence, $I_n$ measures the number of infections per million inhabitants. The eRG equation \eqref{eq:eRG} depends on two constant parameters, $\gamma$ and $A$, which embed different characteristics of an epidemiological wave:
\begin{itemize}
    \item[-] The  $\gamma$ parameter is an effective infection rate, measured in units of $t^{-1}$. It describes how quickly the infectious disease spreads within the population of the region, and it does not depend on the effective number of total infections. As such, $\gamma$ values from different regions can be compared. The numerical value of $\gamma$ encodes all the effects that influence the diffusion speed: the transmissibility of the virus, social and behavioural effects \cite{Cot_2021_google}, and pharmaceutical interventions like vaccinations \cite{cacciapaglia2020better}. All these effects are captured by a single and constant value over the development of a specific wave. 
    \item[-] The  $A$ parameter corresponds to the value of $\alpha$ at the end of the wave, hence it is a measure of the normalised number of infections in the region at the end of the wave. In fact, $A = \alpha (\infty) = \ln I_n (\infty)$. The significance of this parameter is affected by biases in the data collection in each region: for instance, the testing rates and policies. However, as long as these biases remain approximately constant during the development of a wave, the eRG approach can be  effectively applied. 
\end{itemize}
Note that $\gamma$ does not depend on the number of infected individuals. As such, it does not suffer from biases coming from the number of available test kits, nor from testing policies adopted during various phases of the pandemic, nor on any possible regional differences. Hence, $\gamma$ offers a reliable characterisation of the severity of each wave in different regions and at different times.
An advantage of the method lies in the fact that just two parameters ($\gamma$, $A$) suffice to characterise the wave, and they remain practically constant over the evolution of a single wave \cite{Cot_2021_google,cacciapaglia2020better}. The values of $\gamma$ and $A$ can be obtained by fitting the epidemiological data in a specific region with the solution of \eq{eq:eRG}, which is the following logistic function:
\begin{equation}\label{eq:logisticSol}
     \alpha (t) = \frac{A e^{\gamma (t-t_0)}}{1 + e^{\gamma (t-t_0)}}\,,
\end{equation}
where $t_0$ is an integration constant setting the overall timing of the wave. 

In previous works, the eRG framework was extended to include mobility of people among different regions, as long as the flow only involves a small fraction of the region inhabitants \cite{Cacciapaglia:2020mjf}. This extension allowed to study the relation between the emergence of epidemiological waves in different regions, along with the mobility of people among regions. In particular, the timing of the wave peaks could  directly be related to the mobility flow, providing a handle to quantify the impact of various transportation modes on the COVID-19 diffusion.

This feature of the multi-region eRG equations allows us to quantify the impact of the various transportation modes on the diffusion of the infection in the regions of Croatia. The eRG now provides a set of coupled differential equations, one for each $i$-th region under consideration \cite{Cacciapaglia:2020mjf,cacciapaglia2020second}:
\begin{equation} \label{eq:coupledeRG}
    \frac{d \alpha_i}{d t} = \gamma_i\ \alpha_i \left( 1 - \frac{\alpha_i}{A_i} \right) + \sum_{j} \frac{k_{ij}}{N_{m,i}} \left(e^{\alpha_i - \alpha_j} -1 \right)\,, \qquad \alpha_k = \ln \frac{I_k (t)}{N_{m,k}}\,,
\end{equation}
where $k_{ij}$ represents the number of travellers per million inhabitants going from region $i$ to region $j$. The above set of equations also premits the addition of a {\it source region}, i.e. a population of infected individuals that ignites the spread of the infections in the set of regions under consideration \cite{cacciapaglia2020second}. 

In the case of Croatia, the value of the parameters $k_{ij}$ can be estimated by use of the mobility datasets in Table~\ref{tab:01}.

\begin{figure}[t]
\centering
\includegraphics[width=14cm]{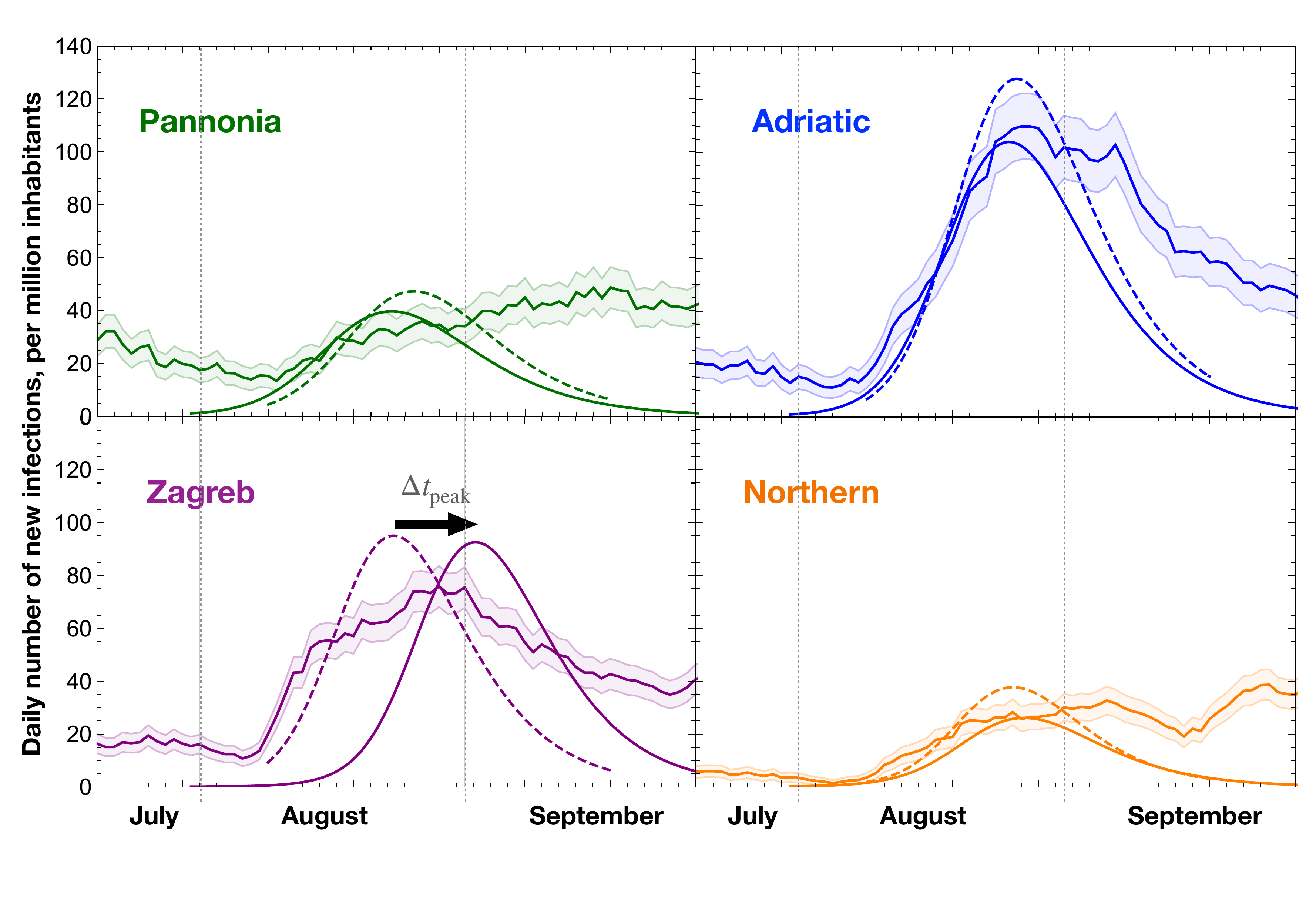}  
\caption{\label{fig:eRG_thirdwave} Comparison of the data (bands with statistical error), eRG modelling of the third wave (dashed curves) and the result of the eRG equations with mobility data (solid curves).  The eRG computation includes all mobility data without relative weight in the number of passengers. The statistical error is computed following a Poissonian distribution on the number of daily new infections, giving an error of $\sigma = \sqrt{N}$ for a count of $N$ new cases. The peak mismatch metric $\Delta t_\mathrm{peak}$ is shown with its orientation for the case of the Zagreb region.}
\end{figure}

To quantify the impact of the mobility datasets listed in Table~\ref{tab:01} on the diffusion of the third wave in Croatia, we adopted the following procedure:
\begin{itemize}
\item	We subdivide Croatia into four regions, chosen to match the Eurostat 2021 NUTS-2 classification.
\item	For each wave, we fit the eRG parameters on the available epidemiological data in each region. For this study, we focus on the second and third waves, occurring between June and September 2020.
\item	We use the eRG equations, together with the fitted parameters, to numerically calculate the diffusion of the third wave to the different regions.
\item As source regions, we use the neighbouring countries (including Italy for the maritime transport) by coupling the eRG equations of the Croatian regions to their epidemiological data. For the flight transportation, we use the epidemiological data of the whole world as a source.  
\item The eRG parameters $\gamma_i$ and $A_i$ are extracted from least-square fits of the Croatian second and third waves. The parameters are listed in \tab{tab:eRG}, where we also indicate the date where the peak occurred, as modelled by the eRG solution.
\begin{table}[]
\centering
\begin{tabular}{ |l|c|c|c|c|c|c| } 
 \cline{2-7}
\multicolumn{1}{c|}{} & \multicolumn{3}{c|}{Second wave} & \multicolumn{3}{c|}{Third wave} \\ \hline
Region & $A$ & $\gamma$  & $t_{\rm peak}$  & $A$ & $\gamma$ & $t_{\rm peak}$\\
 \hline 
 Pannonia & 6.62 & 0.20 & July 4 & 7.01 & 0.13 & August 26 \\ 
 Adriatic & 6.41 & 0.10 & July 15 & 7.93 & 0.14 & August 26 \\ 
 Zagreb & 6.59 & 0.19 & July 2 & 7.58 & 0.15 & August 23 \\ 
 Northern & 5.00 & 0.23 & July 3 & 6.64 & 0.16 & August 26 \\ 
 \hline
\end{tabular}
\caption{Parameters in the eRG solution that we use to model the second and third waves in the four Croatian regions. We also indicate the peak timing $ t_{\rm peak}$ from the eRG modelling of the waves. They correspond to the peaks of the dashed lines in \fig{fig:eRG_thirdwave}.}\label{tab:eRG}
\end{table}
\item We define and study specific scenarios where the different mobility datasets are included with a weight. The latter parameterises the effective impact of the actual transportation modes on the virus diffusion. In practice, this weight corresponds to the probability of finding infected individuals among the passengers of that specific transportation vector.
\item	We compare the result of the numerical equations to the observed data, to establish which configuration offers the best fit to the timing of the third wave in the different regions. As an example, \fig{fig:eRG_thirdwave} shows the third wave obtained by the eRG equations when all transportation passenger are included unweighted. As a quantifier of the model performance, we make use of the shift between the predicted wave peak and that of the epidemiological data, cf. \eq{eq:Deltatpeak} below.
\end{itemize}
Figure~\ref{fig:eRG_thirdwave}  shows that in some regions, like Pannonia and Northern regions, a peak of infections is not readily identified from the data. This is mainly due to the limited statistics available in these regions, while at national level (and in other regions of the world) it was apparent that the COVID-19 diffusion has a wave-like character. Hence, we  identified peaks in the data by using the logistic function suggested by the eRG framework (\eq{eq:logisticSol}), and using the beginning of August as an initial time ($t=0$). The resulting curves are shown in dashed lines in \fig{fig:eRG_thirdwave}, with the $t_{\rm peak}$ listed in \tab{tab:eRG}. Then we compute the shift in the peak prediction from the mobility data as
\begin{equation}\label{eq:Deltatpeak}
\Delta t_{\rm peak} = \left. t_{\rm peak}\right|_{\rm eRG} - t_{\rm peak}\,,
\end{equation}
as indicated in \fig{fig:eRG_thirdwave} for Zagreb. Hence, positive $\Delta t_{\rm peak}$ indicates that the mobility data predicts a delayed peak as compared to the data, while an optimal modelling is achieved if $\Delta t_{\rm peak} \sim 0$.
The numerical computations have been performed on a personal computer, using the  Wolfram’s software {\it Mathematica}.

\section{Results}\label{sec:Results}
 
The results obtained via the eRG framework considers various configurations of the mobility data. In this way, two main research questions could be addressed: the role (if any) of cross border passenger flow via the maritime links and the estimation of the road traffic to and from Zagreb (\sect{sec:sea}).    Both answers are established by studying the individual impact of each transportation mode. Finally, we study the combined effect of all mobility data and determine the optimal configuration to reproduce the epidemiological data (\sect{sec:zagreb}).
In practice, given a set of mobility data, which feeds into the values of $k_{ij}$ in the eRG equations, we compute the timing of the third wave in the four Croatian regions. To quantify the strength of the numerical solution, we compute the time difference between the peak in the solution and the peak observed in the data (obtained via the eRG modelling). We will see what configurations are best suited to represent the epidemiological data, and we will interpret the results in terms of the relevance of the various transportation modes. From previous work \cite{Cacciapaglia:2020mjf}, we know that varying the weight of the passenger number can move the peak time by up to a week, hence empirically we consider an agreement to be good if the location of the peak ($\Delta t_\mathrm{peak}$) is captured within five days.

\begin{figure}[t]
\centering
\includegraphics[width=16cm]{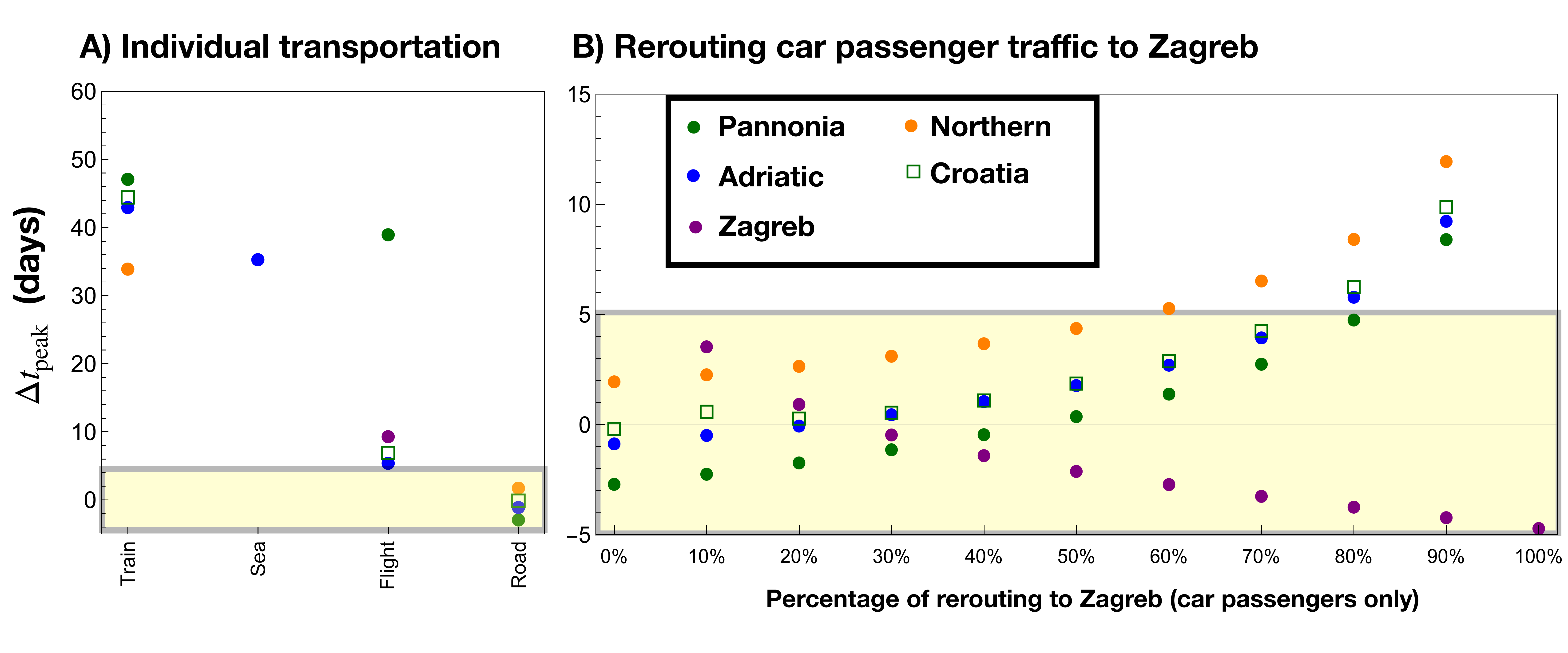}  
\caption{\label{fig:peakmis-rerouting} Results of the eRG computation when including only one transportation mode, with each column corresponding to a different combination of data. The region of acceptable peak mismatch ($\pm 5$ days) is etched in yellow. In Panel A) one transportation mode is included: Train, Sea, Flight and Road data. Missing point correspond to regions where no wave is ignited if just the transportation mode in parenthesis is considered: Pannonia (Sea), Zagreb (Train and Sea), Northern (Sea or Flights).  Flights only marginally reproduce Zagreb and Adriatic, where the main international airports are located, while road transportation  (without rerouting) fully misses Zagreb.
In panel B), road data only is considered, with the indicated percentage of passengers rerouted from all other regions to Zagreb.} 
\end{figure}

\subsection{Single transportation modes and role of maritime mode} \label{sec:sea}

Our aim is to establish what are the main effects of each type of mobility on the infection diffusion. To achieve this, we solved the eRG equations with a single transportation type (“Train”, “Sea”, “Flight”, “Road”) to understand how the data can be reproduced by only using them one at  time. To quantify the fitness of the eRG calculation, in Fig.~\ref{fig:peakmis-rerouting} we show the peak time differences, in days, in each of the four regions, indicated by the solid dots.  Panel ~\ref{fig:peakmis-rerouting}A refers to results obtained including a single dataset, where the missing points indicate regions where a wave start was not triggered by use of just a specific transport mode. 

The results show that none of the transportation modes alone can reproduce the data. In particular, ``Road'' data fails to ignite the wave in Zagreb due to the lack of borders with abroad. Flights play an important role for both Adriatic and Zagreb, where the major international airports are located, while maritime and train transport do not play any significant role in the diffusion of  infections. It is important to highlight that we also report with a square symbol the cumulative data for the whole of Croatia. This was  obtained by solving the eRG equations for Croatia as a single region. These results clearly indicate that both railway and  maritime passengers had a negligible impact on the diffusion of COVID-19 in Croatia.

As a second step, we tested the effect of rerouting a fixed percentage of the road data from the other regions (Adriatic, Pannonia and Northern) to Zagreb. While Adriatic is quite far from Zagreb, we consider that a major road connection to Slovenia (and Italy) goes through the border of the Adriatic region, connecting Zagreb with Rijeka. 

The results are shown in panel~\ref{fig:peakmis-rerouting}B, and are labelled by the percentage of rerouting. It is seen that a rerouting level between 10 and 40\% can well reproduce the epidemiological data for all regions of Croatia ($|\Delta t_\mathrm{peak}|<$ 4 days). This shows that the diffusion of the virus during the third wave in Croatia can be well modelled by using road-only data. In the next section we analyse a more realistic scenario where all transportation modes are included, with the main goal of validating the main conclusions of this analysis.

\subsection{Combined analysis and optimal traffic diversion onto Zagreb} \label{sec:zagreb}

\begin{figure}[t]
\centering
\includegraphics[width=16cm]{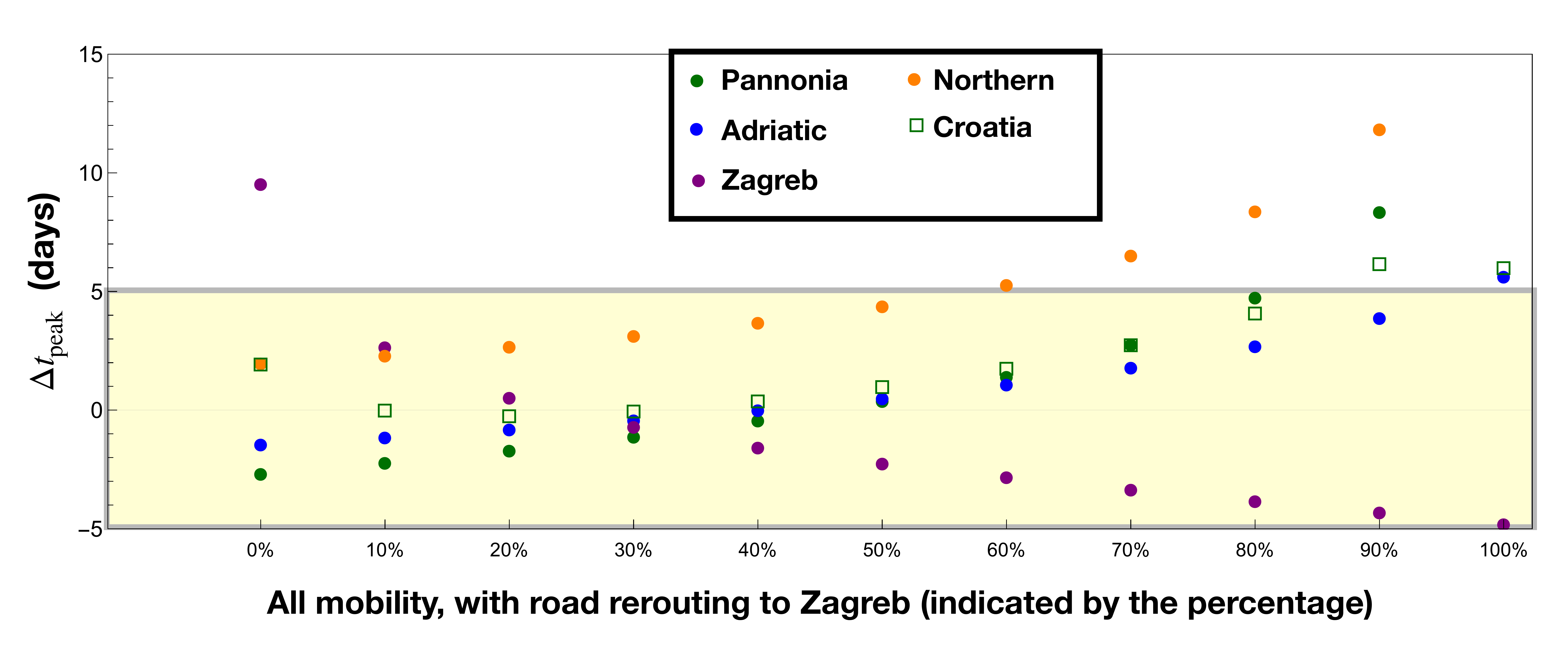}  
\caption{Results including all transportation modes, with a given rerouting from Adriatic, Northern and Pannonia to Zagreb of the road passengers. The inclusion of all transportation passengers improves the results for Pannonia and Adriatic as compared to \fig{fig:peakmis-rerouting}.}\label{fig:peakmis-allmeans} 
\end{figure}

Having investigated the possibility of having a dominant mobility mode for the infection in Croatia, we can clearly see that road traffic is the main factor of virus propagation. As argued in \sect{sec:NUTS},  we had to assume that a  percentage of the  car  traffic going from abroad to Croatia was  directly rerouted to Zagreb.  
We simulated, therefore, a scenario where all mobility data are included, while a certain percentage of the road traffic is rerouted to Zagreb from the Adriatic, Pannonia and Northern regions. This includes the impact of air traffic which can be relevant for both the Adriatic and Zagreb regions. The results are shown in \fig{fig:peakmis-allmeans}. It is still found that a rerouting in the range of 10 -- 40\% can optimally reproduce the epidemiological data in all regions of Croatia, but with smaller mismatches with respect to the results shown in \fig{fig:peakmis-rerouting}B.

We observe a marginal improvement for Pannonia and Adriatic compared to the road-only case, mainly due to the impact of airborne traffic. Considering Pannonia, Adriatic and Zagreb, the best scenario is based on a 30\% road traffic diversion  to Zagreb, where we observe $|\Delta t_\mathrm{peak}|<$ 2 days for those regions.
Note that the Northern region is systematically delayed compared to the peak in the data: this could be due to airborne passengers landing in Zagreb but eventually directed to the Northern region. Another reason may be the poor modelling of the epidemiological data, where a clear peak is not well visible for this region, as shown in \fig{fig:eRG_thirdwave}.

%

\section{Conclusions}\label{sec:Conclusions}

We have quantitatively analysed mobility data along with epidemiological data during the period of June-September of 2020 in Croatia. The eRG framework was used to model the epidemiological data and numerically correlate them to the mobility data. The main goal was to establish the impact of various mobility vectors on the diffusion of the COVID-19 infections, at the origin of the third wave in Croatia between August and September 2020.

Our results show that, although the timing coincided  with the restart of the maritime traffic from Italy after the first lockdown, maritime or train transportation did not play any significant role in the onset of the third wave in August 2020.
Instead, we demonstrated that road mobility was the main contributor, as the car passenger fluxes, when integrated in the eRG framework,  successfully reproduce the timing of the waves in all NUTS-2 regions of Croatia. However, to optimally reproduce the epidemiological data, we assumed that a fraction of the cross-border road passengers were directed to Zagreb, a region which does not have direct borders with neighbouring countries. The inclusion of  airborne passengers yields to optimally matching the data for the Pannonia, Adriatic and Zagreb regions. This means that epidemic wave peaks are reproduced within an error of 3 days, when about 30\% of the car passengers are redirected to Zagreb. The Northern region of Croatia always features a delay in the eRG prediction, limited within the acceptable range of five days.

These results provide a further validation of the eRG method to combine mobility data with a fast and accurate prediction of the next epidemiological wave. However,  due to lack of data, internal mobility within Croatia was not considered. This gap also relates to the cross-border passengers directed to Zagreb while crossing the other regions. Hence, the eRG results could be greatly improved if aggregated internal mobility data were provided, for instance based on smartphone usage and tracking \cite{Cot_2021_google}.

Nevertheless, our study allowed us to deduce a couple of important lessons on the effect of various transportation modes for the diffusion of an infectious disease. First and foremost, the study warns about simplistic association of nearly simultaneous signals during a pandemic. Indeed, different from what expected, no causal link between ferry traffic and onset of the third wave of COVID-19 in Croatia could be assessed. Instead, and this is our second finding, road traffic was found, during those early phases of the pandemic, to be the leading driver of the virus diffusion in Croatia. This  points towards land border control as one of the most effective ways to limit the spread of an infectious disease. However, this approach can be effective only if timely implemented \cite{Cacciapaglia:2020mjf}.
The eRG-based modelling approach in combination with proper mobility datasets can thus provide the means to rule out non-causal relationships, supporting  decision-makers in recognizing the most effective actions at the beginning of a pandemic.


\section*{Acknowledgements}
We acknowledge support from the Croatian Ministry of Maritime Affairs, Transport and Infrastructure (MMPI) in providing CIMIS data. In particular, we are grateful for the work by Mr. Marko Prpić and  Mr. Davor Deželjin (technical experts) as well as by Mr. Tomislav Budić (project team leader at MMPI).  Mrs. Simone Phoré  (CMCC) is thanked for proof-reading the manuscript. 
The whole work was financially supported by the European Regional Development Fund through the Italy-Croatia Interreg programme, project GUTTA, grant number 10043587.

\section*{Author contributions statement}
All authors  participated in the design of this project. DA and AJ collected and analysed mobility data relative to train and car passengers. GC and CC designed the model and the computational framework, while CC provided the numerical results and data analysis. CC, GC and GM led the writing of the manuscript.

\section*{Additional information}
The authors declare no competing interests.

\section*{Data and code availability}
All raw data used in this work are obtained from open source repositories, cited in the main text.

\bibliography{Giacomo,GM_publications,assessments}

\begin{thebibliography}{10}
\expandafter\ifx\csname url\endcsname\relax
  \def\url#1{\texttt{#1}}\fi
\expandafter\ifx\csname urlprefix\endcsname\relax\def\urlprefix{URL }\fi
\expandafter\ifx\csname doiprefix\endcsname\relax\def\doiprefix{DOI }\fi
\providecommand{\bibinfo}[2]{#2}
\providecommand{\eprint}[2][]{\url{#2}}

\bibitem{10.3389/ffgc.2021.626635}
\bibinfo{author}{Plumptre, A.~J.} \emph{et~al.}
\newblock \bibinfo{journal}{\bibinfo{title}{Where might we find ecologically
  intact communities?}}
\newblock {\emph{\JournalTitle{Frontiers in Forests and Global Change}}}
  \textbf{\bibinfo{volume}{4}} (\bibinfo{year}{2021}).
\newblock
  \urlprefix\url{https://www.frontiersin.org/articles/10.3389/ffgc.2021.626635}.
\newblock \doiprefix 10.3389/ffgc.2021.626635.

\bibitem{climatechange}
\bibinfo{author}{Carlson, C.~J.} \emph{et~al.}
\newblock \bibinfo{journal}{\bibinfo{title}{Climate change increases
  cross-species viral transmission risk}}.
\newblock {\emph{\JournalTitle{Nature}}} \textbf{\bibinfo{volume}{607}},
  \bibinfo{pages}{555–562} (\bibinfo{year}{2022}).
\newblock \urlprefix\url{https://www.nature.com/articles/s41586-022-04788-w}.
\newblock \doiprefix 10.1038/s41586-022-04788-w.

\bibitem{permafrost}
\bibinfo{author}{Miner, K.~R.} \emph{et~al.}
\newblock \bibinfo{journal}{\bibinfo{title}{{Emergent biogeochemical risks from
  Arctic permafrost degradation}}}.
\newblock {\emph{\JournalTitle{Nature Climate Change}}}
  \textbf{\bibinfo{volume}{11}}, \bibinfo{pages}{809–819}
  (\bibinfo{year}{2021}).
\newblock \urlprefix\url{https://www.nature.com/articles/s41558-021-01162-y}.
\newblock \doiprefix 10.1038/s41558-021-01162-y.

\bibitem{COVID19-1}
\bibinfo{author}{Zhu, N.} \emph{et~al.}
\newblock \bibinfo{journal}{\bibinfo{title}{{A novel Coronavirus from patients
  with pneumonia in China}}}.
\newblock {\emph{\JournalTitle{N. Engl. J. Med.}}}
  \textbf{\bibinfo{volume}{382}}, \bibinfo{pages}{727–733}
  (\bibinfo{year}{2019}).

\bibitem{COVID19-2}
\bibinfo{author}{Guan, W.} \emph{et~al.}
\newblock \bibinfo{journal}{\bibinfo{title}{{Clinical characteristics of
  Coronavirus disease 2019 in China}}}.
\newblock {\emph{\JournalTitle{N. Engl. J. Med.}}}
  \textbf{\bibinfo{volume}{382}}, \bibinfo{pages}{1708–1720}
  (\bibinfo{year}{2020}).

\bibitem{Bluedot}
\bibinfo{author}{Bogoch, I.~I.} \emph{et~al.}
\newblock \bibinfo{journal}{\bibinfo{title}{{Potential for global spread of a
  novel coronavirus from China}}}.
\newblock {\emph{\JournalTitle{Journal of Travel Medicine}}}
  \textbf{\bibinfo{volume}{27}} (\bibinfo{year}{2020}).
\newblock \urlprefix\url{https://doi.org/10.1093/jtm/taaa011}.
\newblock \doiprefix 10.1093/jtm/taaa011.
\newblock \bibinfo{note}{Taaa011},
  \eprint{https://academic.oup.com/jtm/article-pdf/27/2/taaa011/32902441/taaa011.pdf}.

\bibitem{preparedness1}
\bibinfo{author}{Frutos, R.}, \bibinfo{author}{Gavotte, L.},
  \bibinfo{author}{Serra-Cobo, J.}, \bibinfo{author}{Chen, T.} \&
  \bibinfo{author}{Devaux, C.}
\newblock \bibinfo{journal}{\bibinfo{title}{{COVID-19 and emerging infectious
  diseases: The society is still unprepared for the next pandemic}}}.
\newblock {\emph{\JournalTitle{Environ Res}}} \textbf{\bibinfo{volume}{202}},
  \bibinfo{pages}{111676} (\bibinfo{year}{2021}).
\newblock
  \urlprefix\url{https://www.ncbi.nlm.nih.gov/pmc/articles/PMC8268624/}.
\newblock \doiprefix 10.1016/j.envres.2021.111676.

\bibitem{Lancet}
\bibinfo{author}{Priesemann, V.} \emph{et~al.}
\newblock \bibinfo{journal}{\bibinfo{title}{{Calling for pan-European
  commitment for rapid and sustained reduction in SARS-CoV-2 infections}}}.
\newblock {\emph{\JournalTitle{The Lancet}}} \textbf{\bibinfo{volume}{397}},
  \bibinfo{pages}{92--93} (\bibinfo{year}{2021}).
\newblock \doiprefix 10.1016/S0140-6736(20)32625-8.

\bibitem{Chinazzi}
\bibinfo{author}{Chinazzi, M.} \emph{et~al.}
\newblock \bibinfo{journal}{\bibinfo{title}{{The effect of travel restrictions
  on the spread of the 2019 novel coronavirus (COVID-19) outbreak}}}.
\newblock {\emph{\JournalTitle{Science}}} \textbf{\bibinfo{volume}{368}},
  \bibinfo{pages}{395--400} (\bibinfo{year}{2020}).
\newblock \urlprefix\url{https://science.sciencemag.org/content/368/6489/395}.
\newblock \doiprefix https://doi.org/10.1126/science.aba9757.
\newblock
  \eprint{https://science.sciencemag.org/content/368/6489/395.full.pdf}.

\bibitem{Lai2020}
\bibinfo{author}{Lai, S.} \emph{et~al.}
\newblock \bibinfo{journal}{\bibinfo{title}{{Effect of non-pharmaceutical
  interventions for containing the COVID-19 outbreak in China}}}.
\newblock {\emph{\JournalTitle{Nature}}}  (\bibinfo{year}{2020}).
\newblock \doiprefix https://doi.org/10.1038/s41586-020-2405-7.

\bibitem{Flaxman2020}
\bibinfo{author}{Flaxman, S.} \emph{et~al.}
\newblock \bibinfo{journal}{\bibinfo{title}{{Estimating the effects of
  non-pharmaceutical interventions on COVID-19 in Europe}}}.
\newblock {\emph{\JournalTitle{Nature}}}  (\bibinfo{year}{2020}).
\newblock \doiprefix https://doi.org/10.1038/s41586-020-2293-x.

\bibitem{SEIR}
\bibinfo{author}{Prem, K.} \emph{et~al.}
\newblock \bibinfo{journal}{\bibinfo{title}{{The effect of control strategies
  to reduce social mixing on outcomes of the COVID-19 epidemic in Wuhan, China:
  a modelling study}}}.
\newblock {\emph{\JournalTitle{The Lancet Public Health}}}
  \textbf{\bibinfo{volume}{5, issue 5}}, \bibinfo{pages}{E261 -- E270}
  (\bibinfo{year}{2020}).
\newblock \doiprefix https://doi.org/10.1016/S2468-2667(20)30073-6.

\bibitem{scala2020}
\bibinfo{author}{Scala, A.} \emph{et~al.}
\newblock \bibinfo{journal}{\bibinfo{title}{{Time, space and social
  interactions: exit mechanisms for the Covid-19 epidemics}}}.
\newblock {\emph{\JournalTitle{Sci Rep}}} \textbf{\bibinfo{volume}{10}},
  \bibinfo{pages}{13764} (\bibinfo{year}{2020}).
\newblock \doiprefix https://doi.org/10.1038/s41598-020-70631-9.

\bibitem{Cacciapaglia:2020mjf}
\bibinfo{author}{Cacciapaglia, G.} \& \bibinfo{author}{Sannino, F.}
\newblock \bibinfo{journal}{\bibinfo{title}{{Interplay of social distancing and
  border restrictions for pandemics (COVID-19) via the epidemic Renormalisation
  Group framework}}}.
\newblock {\emph{\JournalTitle{Sci Rep}}} \textbf{\bibinfo{volume}{10}},
  \bibinfo{pages}{15828} (\bibinfo{year}{2020}).
\newblock \doiprefix https://doi.org/10.1038/s41598-020-72175-4.
\newblock \eprint{2005.04956}.

\bibitem{cacciapaglia2020second}
\bibinfo{author}{Cacciapaglia, G.}, \bibinfo{author}{Cot, C.} \&
  \bibinfo{author}{Sannino, F.}
\newblock \bibinfo{journal}{\bibinfo{title}{{Second wave COVID-19 pandemics in
  Europe: A Temporal Playbook}}}.
\newblock {\emph{\JournalTitle{Sci Rep}}} \textbf{\bibinfo{volume}{10}},
  \bibinfo{pages}{15514} (\bibinfo{year}{2020}).
\newblock \doiprefix https://doi.org/10.1038/s41598-020-72611-5.
\newblock \eprint{2007.13100}.

\bibitem{Scudellari}
\bibinfo{author}{Scudellari, M.}
\newblock \bibinfo{journal}{\bibinfo{title}{{How the pandemic might play out in
  2021 and beyond}}}.
\newblock {\emph{\JournalTitle{Nature}}} \textbf{\bibinfo{volume}{584}},
  \bibinfo{pages}{22 -- 25} (\bibinfo{year}{2020}).
\newblock \doiprefix https://doi.org/10.1038/d41586-020-02278-5.

\bibitem{Leskovec2020}
\bibinfo{author}{Chang, S.} \emph{et~al.}
\newblock \bibinfo{journal}{\bibinfo{title}{{Mobility network models of
  COVID-19 explain inequities and inform reopening}}}.
\newblock {\emph{\JournalTitle{Nature}}}  (\bibinfo{year}{2020}).
\newblock \doiprefix https://doi.org/10.1038/s41586-020-2923-3.

\bibitem{March_2021_nature}
\bibinfo{author}{March, D.}, \bibinfo{author}{Metcalfe, K.},
  \bibinfo{author}{Tintor{\'e}, J.} \& \bibinfo{author}{Godley, B.~J.}
\newblock \bibinfo{journal}{\bibinfo{title}{{Tracking the global reduction of
  marine traffic during the COVID-19 pandemic}}}.
\newblock {\emph{\JournalTitle{Nature Communications}}}
  \textbf{\bibinfo{volume}{12}}, \bibinfo{pages}{2415} (\bibinfo{year}{2021}).
\newblock \urlprefix\url{https://doi.org/10.1038/s41467-021-22423-6}.

\bibitem{mannarini_su2022}
\bibinfo{author}{Mannarini, G.}, \bibinfo{author}{Salinas, M.~L.},
  \bibinfo{author}{Carelli, L.} \& \bibinfo{author}{Fass\`{o}, A.}
\newblock \bibinfo{journal}{\bibinfo{title}{{How COVID-19 Affected GHG
  Emissions of Ferries in Europe}}}.
\newblock {\emph{\JournalTitle{Sustainability}}} \textbf{\bibinfo{volume}{14}}
  (\bibinfo{year}{2022}).
\newblock \urlprefix\url{https://doi.org/10.3390/su14095287}.

\bibitem{dundovic2012analysis}
\bibinfo{author}{Dundovi{\'c}, {\v{C}}.}, \bibinfo{author}{Jugovi{\'c}, A.} \&
  \bibinfo{author}{{\v{Z}}galji{\'c}, D.}
\newblock \bibinfo{title}{Analysis of croatian ports in respect to motorways of
  the sea implementation}.
\newblock In \emph{\bibinfo{booktitle}{Proceedings of the 4th international
  maritime science conference}}, \bibinfo{pages}{16--17}
  (\bibinfo{year}{2012}).

\bibitem{cacciapaglia2022earlywarning}
\bibinfo{author}{de~Hoffer, A.} \emph{et~al.}
\newblock \bibinfo{journal}{\bibinfo{title}{{Variant-driven early warning via
  unsupervised machine learning analysis of spike protein mutations for
  COVID-19}}}.
\newblock {\emph{\JournalTitle{Scientific Reports}}}
  \textbf{\bibinfo{volume}{11}}, \bibinfo{pages}{9275} (\bibinfo{year}{2022}).
\newblock \doiprefix https://doi.org/10.1038/s41598-022-12442-8.

\bibitem{webGUTTA}
\bibinfo{title}{{GUTTA project}}.
\newblock \bibinfo{howpublished}{\url{https://bit.ly/guttaproject}}.
\newblock \bibinfo{note}{{Accessed: Feb. 2023}}.

\bibitem{webGUTTAVISIR}
\bibinfo{title}{{GUTTA-VISIR webpage}}.
\newblock \bibinfo{howpublished}{\url{https://www.gutta-visir.eu/}}.
\newblock \bibinfo{note}{{Accessed: Feb. 2023}}.

\bibitem{mannarini_jmse2021}
\bibinfo{author}{Mannarini, G.}, \bibinfo{author}{Carelli, L.},
  \bibinfo{author}{Orovi\'{c}, J.}, \bibinfo{author}{Martinkus, C.~P.} \&
  \bibinfo{author}{Coppini, G.}
\newblock \bibinfo{journal}{\bibinfo{title}{{Towards Least-CO$_2$ Ferry Routes
  in the Adriatic Sea}}}.
\newblock {\emph{\JournalTitle{Journal of Marine Science and Engineering}}}
  \textbf{\bibinfo{volume}{9}} (\bibinfo{year}{2021}).
\newblock \urlprefix\url{https://doi.org/10.3390/jmse9020115}.

\bibitem{Mizumoto_2020}
\bibinfo{author}{Mizumoto, K.} \& \bibinfo{author}{Chowell, G.}
\newblock \bibinfo{journal}{\bibinfo{title}{Transmission potential of the novel
  coronavirus (covid-19) onboard the diamond princess cruises ship, 2020}}.
\newblock {\emph{\JournalTitle{Infectious Disease Modelling}}}
  \textbf{\bibinfo{volume}{5}}, \bibinfo{pages}{264--270}
  (\bibinfo{year}{2020}).
\newblock
  \urlprefix\url{https://www.sciencedirect.com/science/article/pii/S2468042720300063}.

\bibitem{DellaMorte:2020wlc}
\bibinfo{author}{Della~Morte, M.}, \bibinfo{author}{Orlando, D.} \&
  \bibinfo{author}{Sannino, F.}
\newblock \bibinfo{journal}{\bibinfo{title}{{Renormalization Group Approach to
  Pandemics: The COVID-19 Case}}}.
\newblock {\emph{\JournalTitle{Front. in Phys.}}} \textbf{\bibinfo{volume}{8}},
  \bibinfo{pages}{144} (\bibinfo{year}{2020}).
\newblock \doiprefix https://doi.org/10.3389/fphy.2020.00144.

\bibitem{Wilson:1971bg}
\bibinfo{author}{Wilson, K.~G.}
\newblock \bibinfo{journal}{\bibinfo{title}{{Renormalization group and critical
  phenomena. 1. Renormalization group and the Kadanoff scaling picture}}}.
\newblock {\emph{\JournalTitle{Phys. Rev. B}}} \textbf{\bibinfo{volume}{4}},
  \bibinfo{pages}{3174--3183} (\bibinfo{year}{1971}).
\newblock \doiprefix https://doi.org/10.1103/PhysRevB.4.3174.

\bibitem{Wilson:1971dh}
\bibinfo{author}{Wilson, K.~G.}
\newblock \bibinfo{journal}{\bibinfo{title}{{Renormalization group and critical
  phenomena. 2. Phase space cell analysis of critical behavior}}}.
\newblock {\emph{\JournalTitle{Phys. Rev. B}}} \textbf{\bibinfo{volume}{4}},
  \bibinfo{pages}{3184--3205} (\bibinfo{year}{1971}).
\newblock \doiprefix https://doi.org/10.1103/PhysRevB.4.3184.

\bibitem{cacciapaglia2020better}
\bibinfo{author}{Cacciapaglia, G.}, \bibinfo{author}{Cot, C.},
  \bibinfo{author}{Islind, A.~S.}, \bibinfo{author}{{\'O}skarsd{\'o}ttir, M.}
  \& \bibinfo{author}{Sannino, F.}
\newblock \bibinfo{journal}{\bibinfo{title}{{Impact of US vaccination strategy
  on COVID-19 wave dynamics}}}.
\newblock {\emph{\JournalTitle{Scientific Reports}}}
  \textbf{\bibinfo{volume}{11}}, \bibinfo{pages}{10960} (\bibinfo{year}{2021}).
\newblock \doiprefix https://doi.org/10.1038/s41598-021-90539-2.
\newblock \eprint{2012.12004}.

\bibitem{Perc2020}
\bibinfo{author}{Perc, M.}, \bibinfo{author}{Gori\v{s}ek~Miksi\'{c}, N.},
  \bibinfo{author}{Slavinec, M.} \& \bibinfo{author}{Sto\v{z}er, A.}
\newblock \bibinfo{journal}{\bibinfo{title}{{Forecasting COVID-19}}}.
\newblock {\emph{\JournalTitle{Frontiers in Physics}}}
  \textbf{\bibinfo{volume}{8}}, \bibinfo{pages}{127} (\bibinfo{year}{2020}).
\newblock
  \urlprefix\url{https://www.frontiersin.org/article/10.3389/fphy.2020.00127}.
\newblock \doiprefix https://doi.org/10.3389/fphy.2020.00127.

\bibitem{Kermack:1927}
\bibinfo{author}{Kermack, W.~O.}, \bibinfo{author}{McKendrick, A.} \&
  \bibinfo{author}{Walker, G.~T.}
\newblock \bibinfo{journal}{\bibinfo{title}{{A contribution to the mathematical
  theory of epidemics}}}.
\newblock {\emph{\JournalTitle{Proceedings of the Royal Society A}}}
  \textbf{\bibinfo{volume}{115}}, \bibinfo{pages}{700--721}
  (\bibinfo{year}{1927}).
\newblock \doiprefix https://doi.org/10.1098/rspa.1927.0118.
\newblock
  \eprint{https://royalsocietypublishing.org/doi/10.1098/rspa.1927.0118}.

\bibitem{Della_Morte_2021}
\bibinfo{author}{Della~Morte, M.} \& \bibinfo{author}{Sannino, F.}
\newblock \bibinfo{journal}{\bibinfo{title}{{Renormalization Group Approach to
  Pandemics as a Time-Dependent SIR Model}}}.
\newblock {\emph{\JournalTitle{Frontiers in Physics}}}
  \textbf{\bibinfo{volume}{8}} (\bibinfo{year}{2021}).
\newblock \doiprefix https://doi/org/10.3389/fphy.2020.591876.

\bibitem{webNUTS}
\bibinfo{title}{{Eurostat NUTS}}.
\newblock
  \bibinfo{howpublished}{\url{https://ec.europa.eu/eurostat/web/nuts/background}}.
\newblock \bibinfo{note}{{Accessed: Feb. 2023}}.

\bibitem{webHC}
\bibinfo{title}{{Hrvatske Ceste}}.
\newblock \bibinfo{howpublished}{\url{https://hrvatske-ceste.hr/en}}.
\newblock \bibinfo{note}{{Accessed: Feb. 2023}}.

\bibitem{webIRGrail}
\bibinfo{title}{{Independent Regulators' Group IRG-rail 2021 report}}.
\newblock
  \bibinfo{howpublished}{\url{https://www.irg-rail.eu/irg/documents/market-monitoring/312,2021.html}}.
\newblock \bibinfo{note}{{Accessed: Feb. 2023}}.

\bibitem{webKorona}
\bibinfo{title}{{Coronavirus statistics for Croatia, public resource}}.
\newblock \bibinfo{howpublished}{\url{https://www.koronavirus.hr/en}}.
\newblock \bibinfo{note}{{Accessed: Feb. 2023}}.

\bibitem{Cot_2021_google}
\bibinfo{author}{Cot, C.}, \bibinfo{author}{Cacciapaglia, G.} \&
  \bibinfo{author}{Sannino, F.}
\newblock \bibinfo{journal}{\bibinfo{title}{{Mining Google and Apple mobility
  data: temporal anatomy for COVID-19 social distancing}}}.
\newblock {\emph{\JournalTitle{Scientific Reports}}}
  \textbf{\bibinfo{volume}{11}}, \bibinfo{pages}{4150} (\bibinfo{year}{2021}).
\newblock \urlprefix\url{https://doi.org/10.1038/s41598-021-83441-4}.
\newblock \doiprefix 10.1038/s41598-021-83441-4.

\end{thebibliography}

\end{document}